\pdfoutput=1

\documentclass[preprint,12pt]{elsarticle}

\usepackage{amssymb}

\usepackage{amsmath}
\usepackage{svg}

\usepackage{multirow}

\usepackage{caption}
\usepackage{subcaption}

\usepackage{tikz}
\DeclareRobustCommand\full  {\tikz[baseline=-0.6ex]\draw[thick] (0,0)--(0.5,0);}
\DeclareRobustCommand\dotted{\tikz[baseline=-0.6ex]\draw[thick,dotted] (0,0)--(0.54,0);}
\DeclareRobustCommand\dashed{\tikz[baseline=-0.6ex]\draw[thick,dashed] (0,0)--(0.54,0);}

\journal{International Journal of Heat and Mass Transfer}

\begin{document}

\begin{frontmatter}



\title{Pressure-Velocity Coupling in
Transpiration Cooling}


\author{Sophie Hillcoat}
\author{Jean-Pierre Hickey}
\affiliation{organization={Department of Mechanical and Mechatronics Engineering, University of Waterloo},
            addressline={200 University Ave. W.}, 
            city={Waterloo},
            postcode={N2L 3G1}, 
            state={Ontario},
            country={Canada}}

\begin{abstract}

Transpiration cooling is an active thermal protection system of increasing interest in aerospace applications wherein a coolant is effused through a porous wall into a hot external flow. The present work focuses on the interaction between the high-temperature turbulent boundary layer and the pressure-driven coolant flow through the porous wall. Coupling functions were obtained from pore-network simulations to characterize the flow through the porous medium. These were then coupled to direct numerical simulations of a turbulent boundary layer over a massively-cooled flat plate. Two different types of coupling function were used: linear expressions, which do not account for flow interactions between neighbouring pores, and shallow convolutional neural networks (CNN) which incorporate spatial correlations. All coupled cases demonstrated a significant variation in blowing due to the streamwise variation in mean pressure associated with the onset of coolant injection. This trend was reflected in the cooling effectiveness, and was mitigated in the CNN-coupled cases due to the incorporation of lateral flow between neighbouring pores. The distribution of turbulent kinetic energy (TKE) in the coupled cases was also modified by the coupling due to the competing effects of near-wall turbulence attenuation and increased shear due to increasing blowing ratio. Finally, the coupling was shown to impact the power spectral density of the pressure fluctuations at the wall within the transpiration region, attenuating the largest scales of the turbulence whilst leaving the smaller scales relatively unaffected. 

\end{abstract}


\begin{highlights}
\item Direct Numerical Simulations of a turbulent boundary layer with transpiration cooling were conducted, incorporating interactions between the coolant injection velocity and near-wall pressure fluctuations.
\item An algorithm was developed to to couple Direct Numerical simulations with a shallow convolutional neural network to consider the coupling between two flow domains.
\item The importance of lateral flow between neighbouring pores in the porous wall when considering this coupling was demonstrated. Coupling effects were found to be greatly amplified in cases without lateral flow (representative of effusion cooling).
\item Cooling effectiveness is shown to vary significantly in the streamwise direction and the contributions due to the turbulent transport of heat, film accumulation, and heat advection are discussed.
\item The incorporated coupling is found to attenuate the turbulent kinetic energy in the boundary layer flow as well as the largest scale pressure fluctuations at the wall while leaving the smaller scale fluctuations relatively untouched.
\end{highlights}


\end{frontmatter}


\section{Introduction}
\label{sec:intro}
The development of thermal protection systems (TPS) is a critical element in the advancement of aerospace technologies \cite{Uyanna_Najafi_2020}. In applications such as high speed external aerodynamics and propulsion, surfaces are exposed to significant thermal loads that severely limit both the type of materials which can be used and their operational lifetime \cite{esser_innovative_2016, jin_investigation_2023}. Transpiration cooling is a type of active TPS wherein coolant is effused through a porous wall into the main high-temperature flow, where it forms a thermal buffer layer.

Transpiration cooling is of interest due to its increased cooling effectiveness compared to other active TPS such as film and convective cooling \cite{eckert_comparison_1954}. This is due to the two concurrent cooling mechanisms involved: enhanced convective heat transfer within the porous wall, and the formation of a coolant film between the wall and the hot gas. In addition, transpiration cooling has shown promise as a method of drag reduction \cite{jiao_use_2021}, and as a means of controlling laminar-to-turbulent transition \cite{cerminara_turbulence_2023}.

Despite its promise, implementations of transpiration cooling have remained largely limited to research applications due to difficulties obtaining a suitable porous medium and a lack of predictive models. The latter is critical for the design of efficient and reliable transpiration cooling systems, which seek to minimise the coolant flow rate. A reduction in flow rate reduces the momentum of the injected coolant and prevents turbulent mixing in the boundary layer, which impedes coolant film formation. However, it also makes the system more susceptible to the naturally-occuring pressure fluctuations in the boundary layer, which may begin to modulate the flow of coolant and thereby alter the transpiration cooling system performance.


Predictive modelling of transpiration cooling requires the consideration of two tightly coupled flow regions: the porous wall through which the coolant is effused and the external high-temperature flow. The two coupled flow regions have vastly different time and length scales, rendering fully resolved, monolithic simulations computationally intractable at realistic conditions. The works that have taken this approach are limited to low Reynolds number flows and simplified geometries \cite{cerminara_mesoscopic_2020, sharma_numerical_2023, wang_spatial_2022}. It is more common for high-fidelity simulations of transpiration cooling to only model the turbulent boundary layer over the wall and assume a uniform injection of coolant, effectively decoupling any interactions with the porous medium flow \cite{cerminara_turbulence_2023, christopher_dns_2020}.

\begin{figure}[htb]
    \centering
    \includesvg[width=0.9\textwidth]{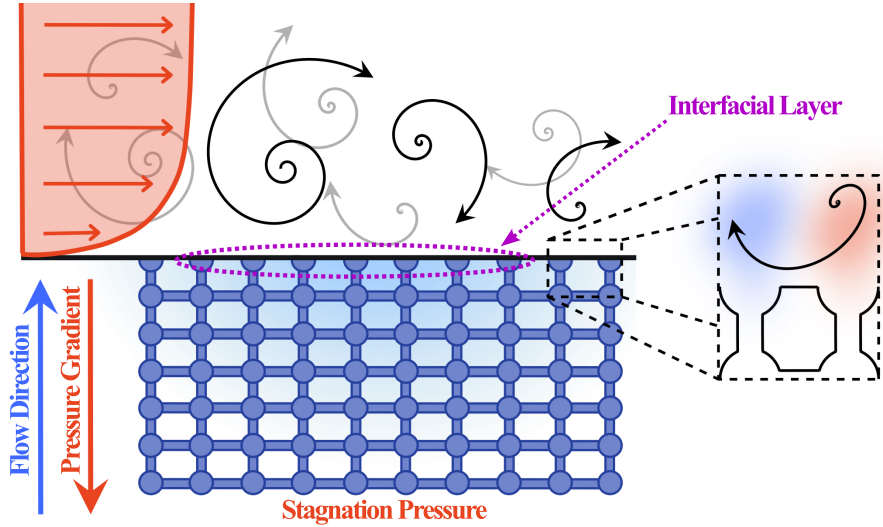}
    \caption{Schematic illustration of the coupling between the flow in the porous medium and the turbulent boundary layer in transpiration cooling.}
    \label{fig:domain_coupling}
\end{figure}

Although the assumption of uniform coolant injection is attractive in its simplicity, in reality it will vary both spatially and temporally due to the aforementioned near-wall pressure fluctuations. The fluctuating pressure at the surface of the porous wall results in local variations in the pressure gradient normal to the porous wall, affecting the local injection mass flow rate, especially at low coolant flow rates. The result is a tight two-way coupling in which pressure fluctuations in the turbulent boundary layer impact the local rate of coolant injection and vice versa, as illustrated in Figure \ref{fig:domain_coupling}.

The vast majority of coupled simulations of transpiration cooling divide the problem into two numerical domains according to the two fluid regions. The transpiration cooling system is then modeled with two segregated solvers that communicate via updated boundary conditions at the interfacial layer (see Figure \ref{fig:domain_coupling}): one for the porous medium flow, solving a Darcy-type equation, and a separate one, solving the Navier-Stokes equations, for the flow above the wall \cite{dahmen_numerical_2014, dahmen_numerical_2017, konig_coupled_2021, prokein_numerical_2021, xiao_large-eddy_2018, zhang_near-wall_2022}. Most of the works that take this approach use Reynolds-Average Navier-Stokes (RANS) solvers to model the main flow, which are neither able to resolve the turbulent structures formed in the boundary layer nor the corresponding pressure fluctuations at the wall. In addition, it has been shown that the typical turbulence closure models in RANS require additional corrections to correctly capture the heat and mass transfer in transpiration cooling \cite{bukva_assessment_2021, chedevergne_transpired_2019}. Furthermore, the continuum modelling approach of the porous medium solvers does not resolve the flow through individual pores and is consequently unable to capture pore-specific flow variations as well as the impact of pore connectivity.

Pore-network modelling (PNM) is an alternative to the continuum approach that is able to resolve the flow at the pore scale while still requiring significantly less resources than a pore-resolved approach. The void region of the material is broken down into disparate pores which are connected via conduits referred to as throats, yielding a pipe network representation of the porous medium. The resulting model preserves information regarding the flow paths through the material and can be used to obtain the mean values of properties such as pressure, temperature, and flow rate at each pore. This method is well-established and has been applied to many different types of problems regarding transport in porous media, including multi-phase transport in fuel cell components \cite{tranter_pore_2016} and gas diffusivity in soil \cite{kiuru_pore_2022}. With respect to Darcy-type single-phase fluid mass transport a system of linear equations is constructed by applying conservation of mass at each pore, which can then be solved using matrix inversion techniques.

The purpose of the present work is explore the dynamics of the coupling between surface pressure fluctuations in the boundary layer and the coolant injection velocity in transpiration cooling. To this end, direct numerical simulations (DNS) of a subsonic turbulent boundary layer over a massively-cooled flatplate were indirectly coupled with PNM simulations of simple porous medium. Two methods are explored for the implementation of the coupling: a linear expression, and a convolutional neural network (CNN). The data obtained from the DNS simulations is used to gain insight into the impact of the explored coupling on coolant distribution and cooling effectiveness as well as the turbulence within the boundary layer.

\section{Numerical Details}
\label{sec:num_details}
\begin{figure}[htb]
    \centering
    \includesvg[width=\textwidth]{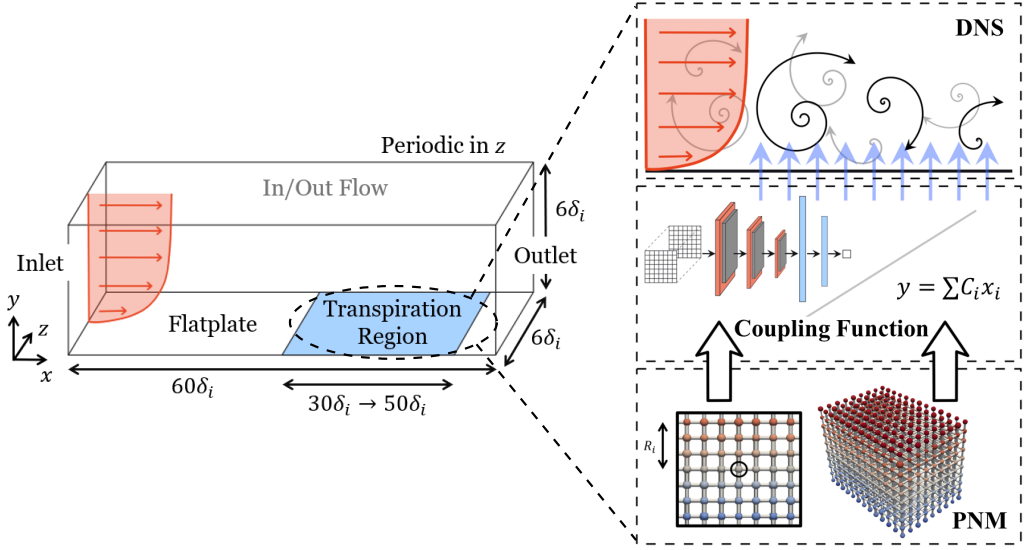}
    \caption{Overview of the computational domain for all DNS cases (left) and the interrelation between the major components in the implemented coupling (right).}
    \label{fig:sketch_numdetails}
\end{figure}

The numerical details of the simulations are divided into three subsections. First, we outline the details of the DNS of a turbulent boundary layer over a cooled, transpirative wall. Second, we discuss the pore-network model used to represent the flow through porous medium. Finally, the details of the coupling functions (neural network and linear expression) are presented. Figure \ref{fig:sketch_numdetails} illustrates the interrelation between these three components.

\subsection{Direct numerical simulation}
\label{sec:numdet_dns}
We use Hybrid, a high-order, compressible DNS solver developed by Bermejo-Moreno et al. \cite{bermejo-moreno_solving_2013}, to simulate the turbulent boundary layer. Hybrid solves the skew-symmetric form of the compressible Navier-Stokes equations on a structured grid. The governing equations are as follow:

\begin{equation}
    \frac{\partial\rho}{\partial t} + \frac{\partial}{\partial x_j}(\rho u_j)=0
    \label{eq:continuity}
\end{equation}

\begin{equation}
    \frac{\partial(\rho u_i)}{\partial t} + \frac{\partial}{\partial x_j}(\rho u_i u_j + p\delta_{ij}) = \frac{\partial\sigma_{ij}}{\partial x_j}
    \label{eq:cons_momentum}
\end{equation}

\begin{equation}
    \frac{\partial(\rho e)}{\partial t} + \frac{\partial[(\rho e + p)u_j]}{\partial x_j} = \frac{\partial (\sigma_{ij}u_i)}{\partial x_j} - \frac{\partial q_j}{\partial x_j}
    \label{eq:cons_energy}
\end{equation}

Here, $\rho$ is the density, $u_i$ is the $i$-th velocity component, $\delta_{ij}$ is the Kronecker delta, $e$ is the total energy, and $q_{ij}$ is the heat flux vector. $\sigma_{ij}$ is the viscous stress tensor:
\begin{equation}
    \sigma_{ij} = 2\mu\left(S_{ij}-\frac{\delta_{ij}}{3}\frac{\partial u_k}{\partial x_k}\right)
\end{equation}

where $S_{ij}$ is the rate-of-strain tensor. The above equations are used in conjunction with the ideal gas equation $p=\rho RT$ to solve for the five conserved variables: $\rho$, $\rho u$, $\rho v$, $\rho w$, and $\rho e$. Aliasing errors are reduced by using a split form of the convective terms and the spatial derivatives are approximated using a sixth-order central finite-difference scheme with a higher-order spatial filtering to stabilise the numerics. Time integration is done using a fourth-order Runge-Kutta scheme.

In addition to its accuracy, speed, and parallelisability, Hybrid was chosen as it was previously modified by Christopher et al. \cite{christopher_dns_2020} for the transpiration cooling case with uniform coolant injection.  As in this previous work, a single fluid is used for both the hot gas and coolant, and the temperature dependence of the fluid viscosity is modeled using a power-law function:

\begin{equation}
    \mu = \mu_0 \left(\frac{T}{T_0}\right)^{0.75}
\end{equation}

Both $\mu_0$ ($1.47\cdot10^{-4}$) and $T_0$ (0.5) are user-defined constant reference values based on the wall conditions. Non-dimensional units are used such that the inlet boundary layer thickness $\delta_i$ and the free-stream density, streamwise velocity, and temperature ($\rho_\infty$, $U_\infty$, $T_\infty$) are all unity. The Prandtl number is fixed (0.7), and the Mach number, based on the freestream velocity, is set to 0.3 to minimise the effects of compressibility.

The flow domain is shown in Figure \ref{fig:sketch_numdetails} and has a size of $60\times6\times6$ with respect to the inlet boundary layer thickness. The mean flow profile and Reynolds stresses at the inlet are taken from those of an established DNS of an incompressible turbulent boundary layer with $Re_\tau=450$ \cite{jimenez_turbulent_2010}. The boundary condition at the inlet plane is set by superimposing the mean flow profile with synthetically generated turbulence corresponding to the imposed Reynolds stress profiles. The resulting signal is digitally filtered in both space and time to ensure a realistic structure of the generated turbulence fluctuations. In/outflow boundary conditions are assigned to both the top and outlet of the flow domain using a summation-by-parts scheme with a simultaneous approximation term penalty, with sponge layers to dampen any numerical reflections. Periodic boundary conditions are applied at the front and back faces (spanwise). 

The wall at the bottom face of the domain is set to a uniform temperature of $T_w=0.5T_\infty$ and is split into two segments. At $0\leq x < 30$ and $x\geq50$ it corresponds to a regular non-permeable, no-slip wall boundary condition. However, for $30\leq x < 50$ a modified no-slip boundary condition is used to represent transpiration wherein a non-zero wall-normal is permitted. Depending on the case, the magnitude of this velocity is either a prescribed constant (uniform blowing) or is set by the coupling function each time step based on the surface pressure distribution.

The degree of transpiration is quantified using the blowing ratio $F=\rho_c v_c/\rho_\infty u_\infty$ where $v_c$ is the Darcy velocity of the coolant (volume of fluid per unit time through a unit cross-section of the porous medium). In all instances, a mean blowing ratio of $\overline{F}=0.6\%$ was targeted over the transpiration region, although in the coupled cases there is some variation in the local blowing ratio due to changes in pressure across the transpiration region.

In all cases, the same fully-structured mesh with a wall-normal grid refinement was used for the DNS, containing roughly 118 million grid points (N$x$, N$y$, N$z$ = 2560, 180, 256). All of the coupled cases were started from the uniform blowing case, which had been run for over 10 mean eddy turnover times and reached a statistically steady state. They were then allowed to run for a minimum of four additional mean eddy turnover times with the implemented coupling before collecting averages. Averages were then collected over a minimum of two eddy turnover times and include averaging in the spanwise direction.

\subsection{Pore-network model}
\label{sec:pnm_details}
The Python package OpenPNM \cite{gostick_openpnm_2016} was used to simulate flow through a simple porous medium based on a cubic lattice structure. The PNM solver uses the Hagen-Poiseuille equation \eqref{eq:hagen-poiseuille}, which is used to estimate the Darcy flow between two connecting pores:

\begin{equation}
    q_{i\rightarrow j} = -\frac{A_{eff}^2}{8\pi\mu L}\Delta p_{i\rightarrow j} = -g_{i\rightarrow j}\Delta p_{i\rightarrow j}
    \label{eq:hagen-poiseuille}
\end{equation}

where $q_{i\rightarrow j}$ is the volumetric flow rate through the throat connecting pores $i$ and $j$, $\Delta p_{i\rightarrow j}$ and $g_{i\rightarrow j}$ are the associated pressure drop and hydraulic conductivity, respectively, and $A_{eff}$ is the conduit effective cross-sectional area and depends on the cross-sectional areas of pores $i$ and $j$ and that of the throat connecting them. Conservation of mass at each pore is then expressed as follows:

\begin{equation}
    \sum_{j=1}^N{g_{i\rightarrow j}\Delta p_{i\rightarrow j}} = q_i = 0
    \label{eq:pnm_cons_momentum}
\end{equation}

\begin{figure}[ht]
    \centering
    \includegraphics[width=0.6\textwidth]{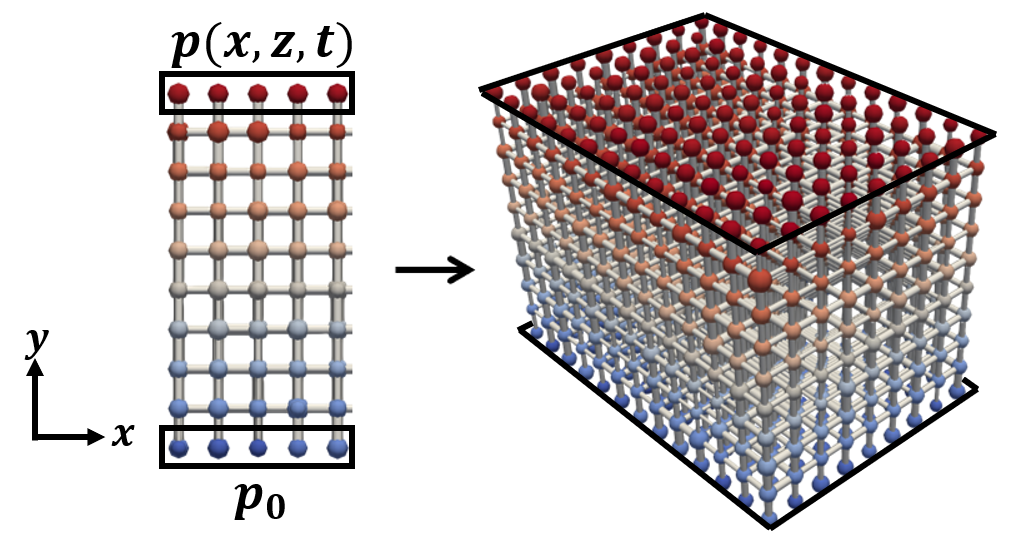}
    \caption{Straight-on and isometric views of a small pore network, showing the basic structure and relevant boundary conditions.}
    \label{fig:pores_bothviews}
\end{figure}

Herein we consider a pore network composed of 2.2 million spherical pores ($200\times43\times256$) connected by cylindrical throats. These dimensions were chosen such that the wall-normal thickness corresponds to the inlet boundary layer thickness from the DNS. The basic structure is illustrated in Figure \ref{fig:pores_bothviews}, which shows a small subsets of the full pore network. The diameter of the pores is randomised according to a normal distribution in order to obtain the desired porosity. Two different pore-networks were constructed according to the details described in Table \ref{tab:pnm_nets} and indirectly coupled to the DNS to explore the impact of the pore geometry on the coupling.

\begin{table}[ht]
    \centering
    \begin{tabular}{|c|c|c|c|c|}
    \hline
        $\boldsymbol{\phi}$ & $\boldsymbol{K\cdot10^7}$ & $\boldsymbol{\overline{d_p}}$ & $\boldsymbol{\sigma_d}$ & $\boldsymbol{\Delta x_i}$ \\
    \hline\hline
        $24.6\%$ & 2.82 & 0.01150 & 0.0010 & \multirow{2}*{0.0234375}\\
        \cline{1-4}
        $30.7\%$ & 4.75 & 0.01295 & 0.0011 & \\ 
    \hline
    \end{tabular}
    \caption{Details of the constructed pore networks coupled with a DNS.}
    \label{tab:pnm_nets}
\end{table}

It is important to note that all cases have the same centre-to-centre distance between neighbouring pores along all three axes ($\Delta x_i$), which corresponds to the streamwise and spanwise grid spacing of the DNS grid at the transpirative wall. This was required for the network to be compatible with the coupling method. Similarly, viscosity and temperature were taken from the conditions set in the DNS at the transpirative wall and all variables are solved for in the same non-dimensional units. 

The top layer of pores in the network correspond to those in the interfacial layer defined in Figure \ref{fig:domain_coupling}. A spatially and temporally varying pressure signal $p(x,z,t)$ is applied at these pores as shown in Figure \ref{fig:pores_bothviews}, whereas a constant pressure boundary condition $p_0$ is applied at the pores on the opposite (bottom) face. The value of $p_0$ is set to obtain a desired mean blowing ratio of 0.6\%, corresponding with the DNS cases.

Two choices of the interfacial layer pressure signal $p(x,z,t)$ were selected and PNM simulations of both networks were run for each one. In the first, the pressure signal corresponds to the wall-pressure signal extracted from the uniform blowing DNS case. The second signal is composed of values randomly selected from a normal distribution with the same mean and standard deviation as the signal extracted from the uniform blowing DNS. The linear regressions and CNN training were performed on the PNM simulations using the latter signal to ensure that they were not biased towards particular spatial correlations. The PNM simulations using the DNS pressure signal were then used to evaluation the performance of the resulting coupling functions, as described in Section \ref{sec:function_performance}.

All PNM simulations were run for over 1000 time steps. It should be noted that in its form presented here PNM is unable to capture transient effects of the flow. Due to the low blowing ratio and fixed wall temperature in the considered cases, it is assumed that the flow in the porous medium is laminar, isothermal, and incompressible. Consequently, information propagation is expected to be \textit{nearly} instantaneous, allowing for the use of steady-state PNM computations.

\subsection{Coupling functions}
Coupling functions were used to connect the PNM simulations in the porous wall with the DNS of the turbulent boundary layer through the interfacial boundary of the transpiration region. This setup allows for a computationally tractable coupling of the flow through a large-scale representative porous wall and a high-fidelity boundary layer simulation.  The coupling function takes the local pore conductivity and pressure at the porous wall as input and outputs the predicted local flow rate. This is then used as the inflow boundary in the DNS. Two forms of coupling function are considered: a linear expression and a neural network. 

\subsubsection{Linear expression}
\label{sec:numdet_linexp}
The linear coupling function expresses the coolant injection velocity at an interfacial pore as a linear function of the local pressure drop across the wall ($\Delta p_{ij}$) and the hydraulic conductivity of the conduit connecting this pore to the bulk porous network ($g_{ij}$). Importantly, this method does not consider the influence of neighbouring pores in the network via the redirection of the internal flow.

The use of this linear expression approach is motivated by the well-established Darcy law, which states that the bulk Darcy velocity and the wall-normal pressure gradient have a linear relationship. Being a bulk medium approach, it is not suitable for capturing local velocity variations. However, it provides a good benchmark against which to compare the performance of the more sophisticated neural network approach.

Rather than defining a single constant to relate the pressure and injection velocity in Hagen-Poiseuille-type equation (such as in the work of Jim\'enez et al. \cite{jimenez_turbulent_2001}), an alternative expression was found by fitting the PNM results with the random pressure signal boundary condition using a multi-variable linear regression:


\begin{equation}
    v_c(g_{ij},\Delta p_{ij}) = C_1 + C_2g_{ij} + C_3\Delta p_{ij} + C_4g_{ij}\Delta p_{ij}
\end{equation}

 $C_1, C_2, C_3, C_4$ are constants obtained from the linear regression and are unique to each of the pore networks considered. This was done in an attempt to extend the bulk medium formulation to incorporate local variations in the pore-network and improve the accuracy of the fit.

\subsubsection{Convolutional neural network}
The linear expression coupling function described above is inherently limited because it only considers the immediate wall-normal pressure gradient and therefore cannot incorporate lateral flow between neighbouring pores. To account for this effect, a shallow CNN was developed using PyTorch \cite{ansel_pytorch_2024}. A CNN was chosen as it is particularly well-suited to problems that require accounting for spatial correlations in data.

The neural network takes as input two ``images" consisting of the pressure gradient across the porous medium and the hydraulic conductivity of the conduits connecting the interfacial pores to the main pore network. This data is then passed through a sequence of convolution and rectified linear unit (ReLU) activation layers followed by fully-connected linear layers. The final output of the network is a single scalar value representing the wall-normal coolant injection velocity at the central pore in the images.

\begin{figure}[ht]
    \centering
\includegraphics[width=0.6\textwidth]{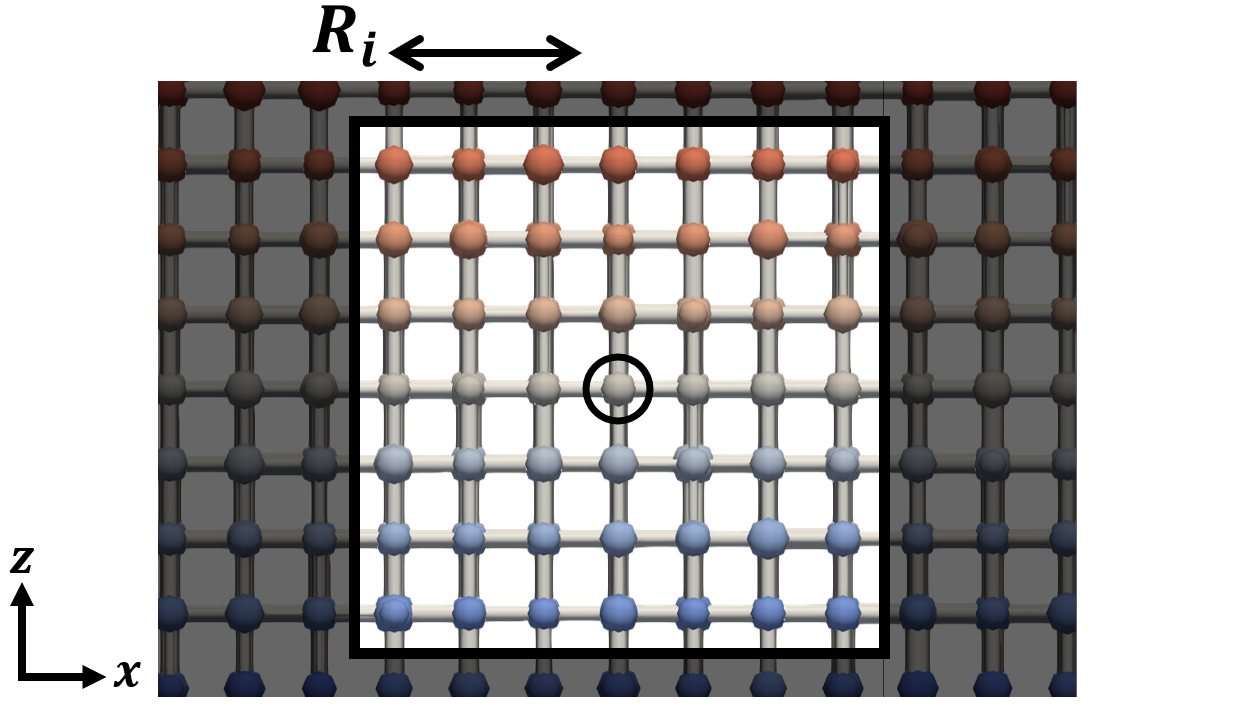}
    \caption{Illustration (top view) of the region captured by the radius of influence $R_i$ around the circled central pore.}
    \label{fig:Ri_sketch}
\end{figure}

To keep the network small, it is necessary to limit the size of the field used as input. Therefore, a radius of influence $R_i$ is defined that dictates the maximum distance from the central pore to include in the field in terms of number of pores. The actual field given to the CNN is then defined by a square whose inscribed circle has a radius of $R_i$ (see Figure \ref{fig:Ri_sketch}).

Upon inspection of the cross-correlation between the coolant injection velocity and surface pressure on the wall a radius of influence of three pores was selected for the present cases. The impact of this selection is discussed further in Section \ref{sec:function_performance}.

\begin{figure}[ht]
    \centering
    \includegraphics[width=0.8\textwidth]{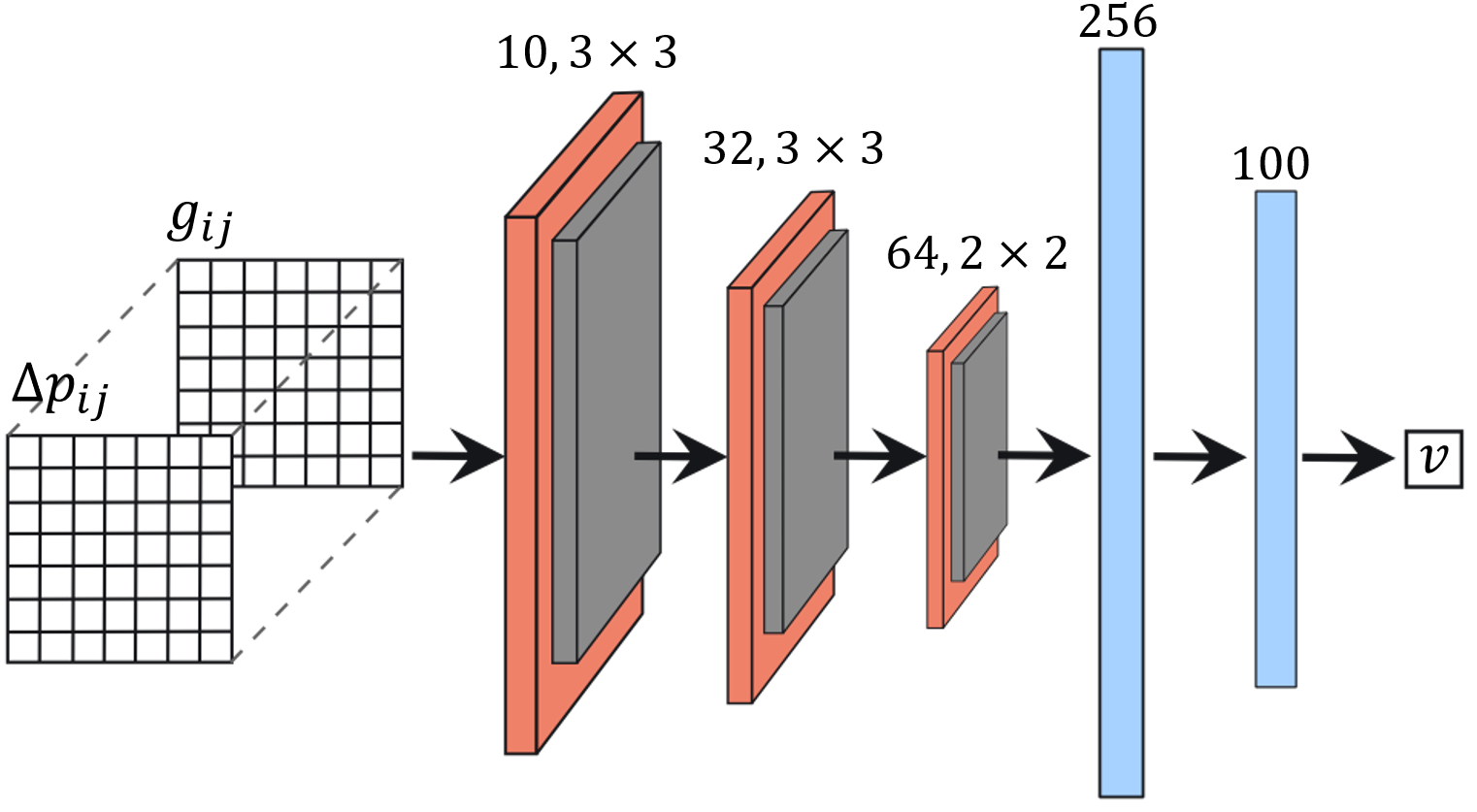}
    \caption{Sketch of the basic structure of the neural network which was coupled with the DNS. The red represents convolution layers of varying channel and kernel sizes, the grey ReLU activation layers, and the blue indicates fully-connected linear layers.}
    \label{fig:nn_structure}
\end{figure}

A sketch of the final structure of the neural network is given in Figure \ref{fig:nn_structure}. It takes as input two $7\times7$ images containing the pressure drop across the wall and the surface pore hydraulic conductivity in the area surrounding the central pore of interest. These are passed through three 2D convolution layers, the first two with a kernel size of $3\times3$ and the last with a kernel size of $2\times2$. The number of channels is 10, 32, and 64, respectively. In all cases a unit stride is used, no padding is applied, and a ReLU activation layer is applied following each convolution. The subsequent output is flattened and passed through two fully-connected layers of size 256 and 100, respectively, to yield the coolant injection velocity at the central pore of interest. 

Two neural networks were trained, corresponding to the two different pore networks. In both cases, the Adam optimiser was used for training in conjunction with a root mean square (RMS) error loss function. In each case, the network was allowed to train until the loss function plateaued, corresponding to roughly 50 epochs. The learning rate was initially left at $10^{-3}$ and was reduced up to $10^{-6}$ as the training progressed.

\subsection{Simulated cases}
Two DNS cases were simulated for each pore network, for a total of four distinct cases: two coupled using a linear expression, and two coupled using a CNN. Two additional cases with the same DNS setup, corresponding to the cases without coolant injection and with uniform coolant injection, were taken from the work done by Christopher et al. \cite{christopher_dns_2020}. A summary of the relevant cases is given in Table \ref{tab:cases}. The PNM simulations described in Section \ref{sec:pnm_details} are not included; they were used to obtain the coupling functions that are used herein. The porosities of the pore networks are in the low-to-mid range of those typically chosen for transpiration systems (between 10\% and 40\%) \cite{wang_performance_2004, wang_experimental_2014, wu_optimization_2018, bohrk_transpiration_2015}.

\begin{table}[htbp]
    \centering
    \begin{tabular}{|l|c|c|}
    \hline
        \textbf{Coupling} & \textbf{Case} & \textbf{Notes} \\
    \hline\hline
        \multirow{2}*{None} & No Blowing & From Christopher et al. \cite{christopher_dns_2020} \\
        & Uniform Blowing & From Christopher et al. \cite{christopher_dns_2020} \\
        \hline
        \multirow{2}*{Linear Regression} & LR-24 & $\phi=24.6\%$\\
        & LR-30 & $\phi=30.7\%$\\
        \hline
        \multirow{2}*{Neural Network} & NN-24 & $\phi=24.6\%$\\
        & NN-30 & $\phi=30.7\%$\\
        
    \hline
    \end{tabular}
    \caption{Summary of the DNS cases.}
    \label{tab:cases}
\end{table}

\section{Results}
\label{sec:results}
For the sake of brevity, the results presented herein focus on the pore network with the higher porosity ($\phi=30.7\%$). The changes due to the coupling are best accentuated in the results corresponding to this network due to its increased sensitivity to changes in the wall-normal pressure gradient.

\subsection{Coupling function performance}
\label{sec:function_performance}
\begin{figure}[htbp]
    \centering
    \begin{subfigure}[b]{0.425\textwidth}
        \centering
        \includegraphics[width=\textwidth]{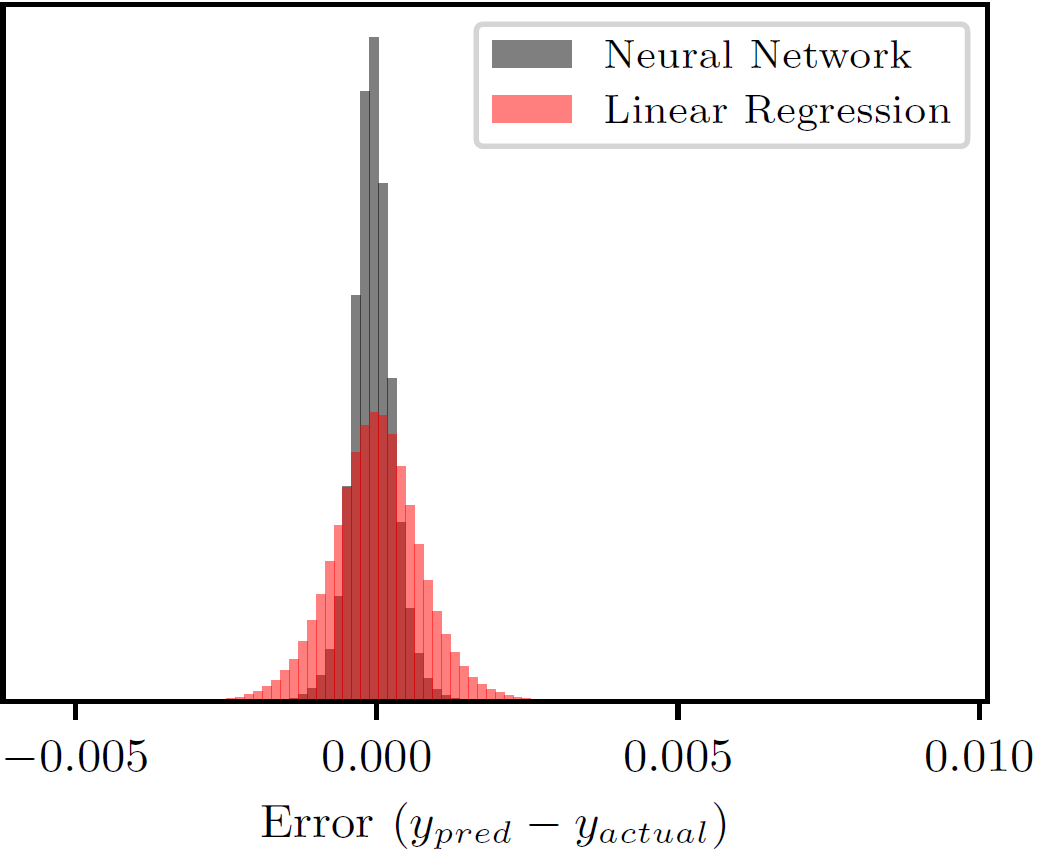}
        \caption{Random interfacial pressure signal (no coherent structures).}
        \label{fig:errhist_random}
    \end{subfigure}
    \hfill
    \begin{subfigure}[b]{0.515\textwidth}
        \centering
        \includegraphics[width=\textwidth]{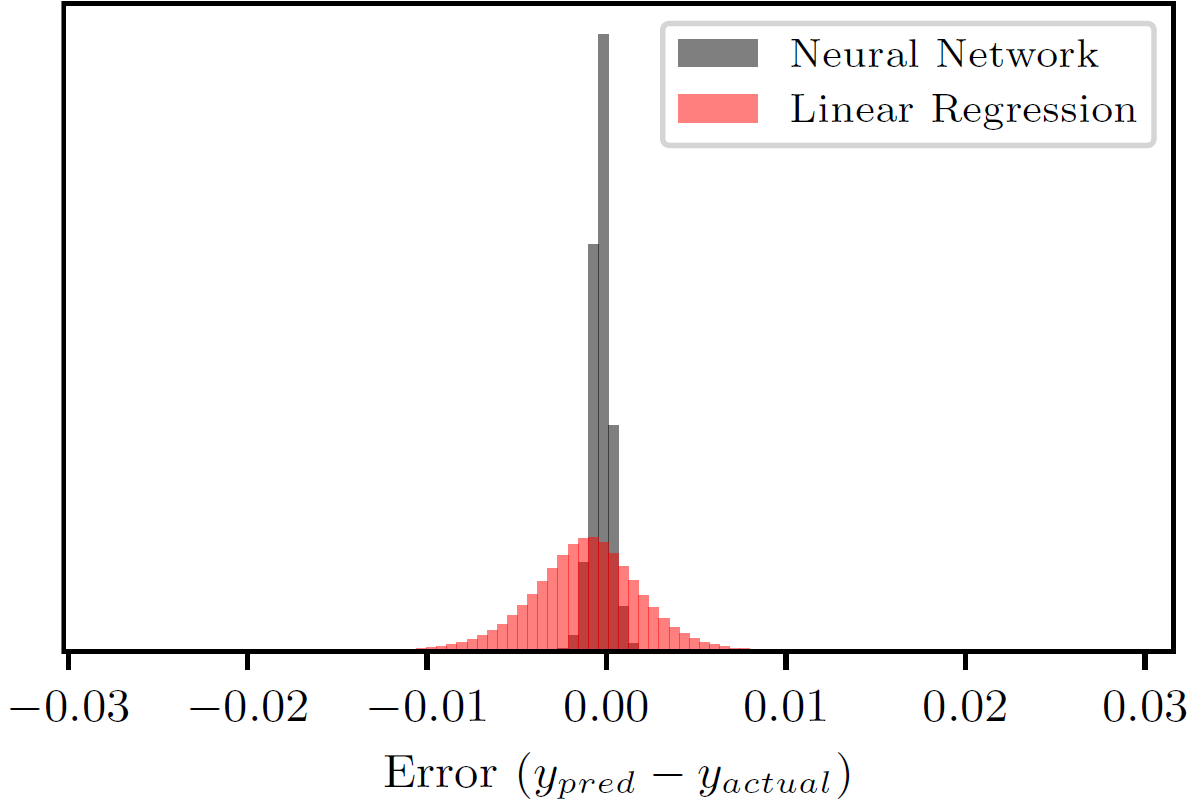}
        \caption{Interfacial pressure signal extracted from the uniform blowing DNS}
        \label{fig:errhist_dns}
    \end{subfigure}
    \caption{Error histograms of both the CNN (grey) and linear regression expression (red) when applied to the PNM simulations.}
    \label{fig:coupling_errhist}
\end{figure}

The implementation of the present indirect coupling relies on the ability of the chosen coupling functions to accurately capture the behaviour of the flow within in the porous medium. The error associated with the coupling functions is quantified as the difference between the local injection velocity predicted by the coupling function and the corresponding value extracted from the PNM simulations. The error histograms produced when applying both coupling functions to the PNM simulations are shown in Figure \ref{fig:coupling_errhist}. For the present cases, the mean injection velocity corresponding to the selected mean blowing ratio is $\overline{v_c}=0.003$.

While both coupling functions perform similarly when applied to the PNM simulation with the random pressure signal, the CNN significantly outperforms the linear expression when both are applied to the simulation using the pressure signal extracted from the uniform blowing DNS. The extracted pressure has natural spatial correlations due to the turbulent structures at the wall. Consequently, there are coherent pressure ``footprints" at the wall which cover multiple surface pores and are able to strongly influence the flow inside the porous medium. 

The linear expression, which cannot account for the influence of flow between neighbouring pores, is unable to capture this effect, resulting in the large errors observed. It is therefore not suitable for implementing the coupling effects between the two flow domains in transpiration cooling, being more representative of an effusion cooling-type scenario where there is no lateral flow of coolant prior to injection. In contrast, the increase in the error of the predictions made by the CNN is much more moderate due to its ability to incorporate these spatial interactions. This highlights the importance of considering interactions between nearby pores when modelling the coupling of the porous medium domain with the main turbulent flow in transpiration cooling.

\begin{figure}[htb]
    \centering
    \includegraphics[width=0.7\textwidth]{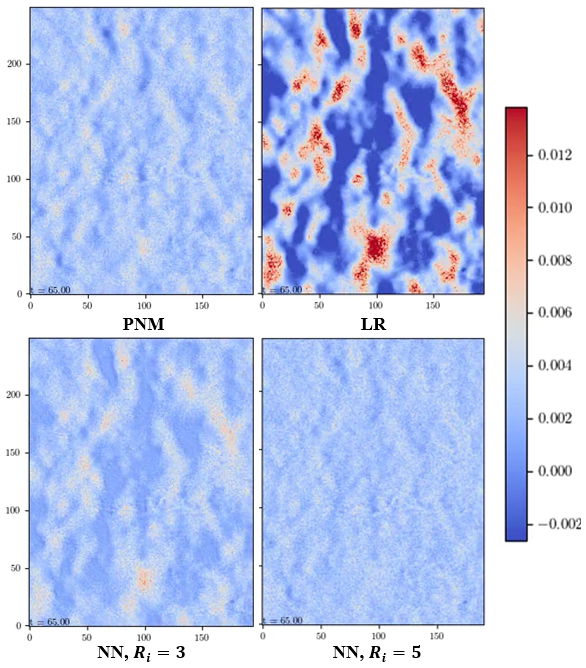}
    \caption{Instantaneous wall-normal injection velocity as given by the PNM simulation using the extracted pressure signal (top left) and predicted by the linear regression expression (top right), CNN with $R_i=3$ (bottom left), and CNN with $R_i=5$ (bottom right).}
    \label{fig:nn_lr_vpnm_comparison}
\end{figure}

It should be noted that, despite being an improvement over those of the linear expression, the errors in the CNN predictions are not insignificant. One major source of the observed errors is the selected radius of influence $R_i$ which limits the size of the field surrounding the central pore that is considered. Figure \ref{fig:nn_lr_vpnm_comparison} shows instantaneous snapshots of the injection velocity field from the PNM simulation with the extracted pressure signal, as well as those predicted by the linear regression and two CNNs with different values of $R_i$.

As previously established, the predictions obtained using linear regression significantly deviate from the PNM results. It is also apparent that these errors affect the overall structure of the signal, further reinforcing the previous conclusion that it is unable to capture the interactions of neighbouring pores. What is of greater interest, however, is the difference between the predictions of the two neural networks, which demonstrates the impact of the selected radius of influence.

The CNN with the smaller radius of influence of three is the same as the one used to generate the error histograms in Figure \ref{fig:coupling_errhist}. Although it is able to recreate the general structures and trends from the PNM case, it also has the effect of smoothing out the smaller visible structures. Meanwhile, the neural network with an increased radius of influence of five is able to capture these small structures, which is reflected in a corresponding reduction in the RMS error. However, it underpredicts the magnitude of the deviations in the injection velocity from the mean.

While the neural network with the larger radius of influence does perform better overall, the CNN with a radius of influence of three was nonetheless used for the coupled DNS cases. This is because increasing the radius of influence is associated with a significant increase in the computational expense of the CNN, and pore-network modelling itself comes with important limitations that make it not strictly applicable for the present work. In light of this, it was decided to sacrifice some accuracy in order to have a neural network that is lightweight enough to couple with a DNS. This network is still able to reproduce the general reaction of the local coolant injection to changes in the surface pressure, and is therefore sufficient for the present purpose.

\subsection{Injection Distribution}
\label{sec:injection_dist}
\begin{figure}[htb]
    \centering
    \includegraphics[width=\textwidth]{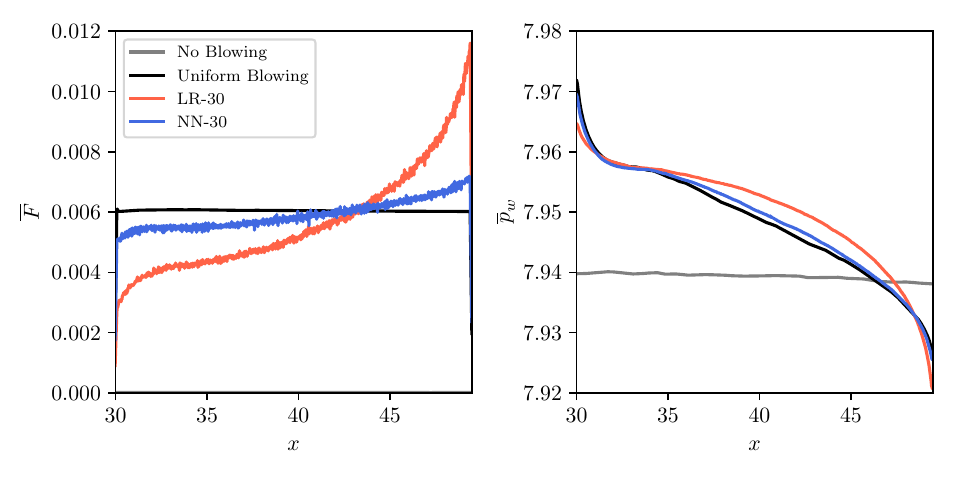}
    \caption{Streamwise variation in blowing ratio (left) and mean wall pressure (right). The blowing ratio $\overline{F}$ for the no blowing case is present but not visible as it is directly on the x-axis.}
    \label{fig:mean_FP}
\end{figure}

One of the simplest metrics used to evaluate the influence of the imposed coupling is the streamwise distribution of coolant injection. Changes in the local rate at which coolant is effused through the porous wall reflect changes to the local flow structure and transpiration performance parameters. This is quantified by the streamwise blowing ratio, which is shown alongside the corresponding mean pressure at the wall in Figure \ref{fig:mean_FP}. 

Both the linear expression and neural network couplings have a significant impact on the injection of cooling, with a reduced blowing ratio at the beginning of the transpiration region relative to the uniform blowing case. The blowing ratio increases further downstream, eventually exceeding that of the uniform blowing case. 

The streamwise increase in the blowing ratio corresponds directly to a decrease in the mean pressure at the wall through the transpiration region. This is expected; a larger pressure at the wall results in a reduced wall-normal pressure gradient across the porous medium, which locally decreases the coolant injection velocity. The mean pressure over the transpiration region, in all three cases with coolant injection, is notably larger than the no blowing case. This is due to the boundary layer flow flow suddenly encountering the wall-normal coolant injection at the beginning of the transpiration region. The pressure eventually decreases as the flow adapts to the change, and continues to decrease as the accumulation of the coolant film lifts the turbulent flow off of the wall.

With respect to the two different coupling functions employed, LR-30 demonstrates a significantly increased sensitivity to pressure variations relative to NN-30, suggesting that incorporating the influence of flow between neighbouring pores within the porous medium reduces the impact of the coupling.

\subsection{Cooling effectiveness}
\label{sec:results-eta}
The cooling effectiveness is one of the most important parameters to consider when it comes to transpiration cooling. It can be computed using the ratio of the wall heat flux of the considered case to that of the reference case without coolant injection:

\begin{equation}
    \eta=1-\frac{q_w}{q_{w,0}}=1-\frac{\left.(\partial T/\partial y)\right|_w}{\left.(\partial T/\partial y)\right|_{w,0}}
\end{equation}

\begin{figure}[htb]
    \centering
    \includegraphics[width=\textwidth]{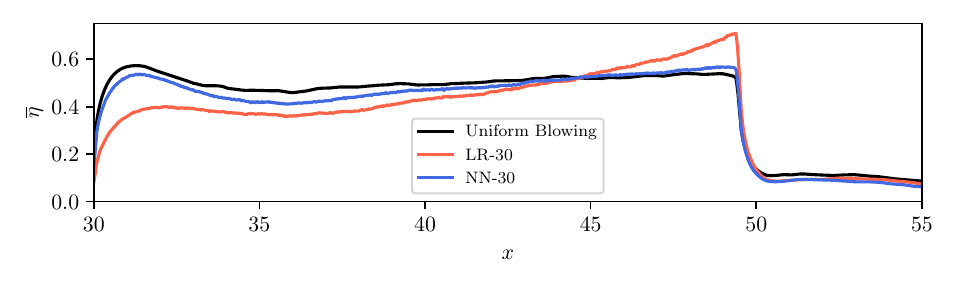}
    \caption{Mean cooling effectiveness along the spanwise direction.}
    \label{fig:mean_eta}
\end{figure}

The time and spanwise-averaged cooling effectiveness along the streamwise direction was calculated for each of the DNS cases, as shown in Figure \ref{fig:mean_eta}. Overall, the same trends are visible as were observed previously when looking at the streamwise variation in blowing ratio, which is expected as the two are closely tied. To gain a deeper understanding, it is necessary to looking at the relative importance of the compononents contributing to cooling effectiveness.

\subsubsection{Turbulent heat flux}
\label{sec:turbheat_transport}
One notable difference incurred by the coupling is a smoothing out of the peak in cooling effectiveness at the beginning of the transpiration region. This peak was previously explained by Christopher et al. \cite{christopher_dns_2020} by considering the delay in the development of vortical structures produced due to the onset of coolant injection. This delay initially mitigates the increase in turbulent transport of heat to the wall, creating a peak in cooling effectiveness until the produced vortices develop sufficiently farther downstream.

\begin{figure}[htb]
    \centering
    \includegraphics[width=\textwidth]{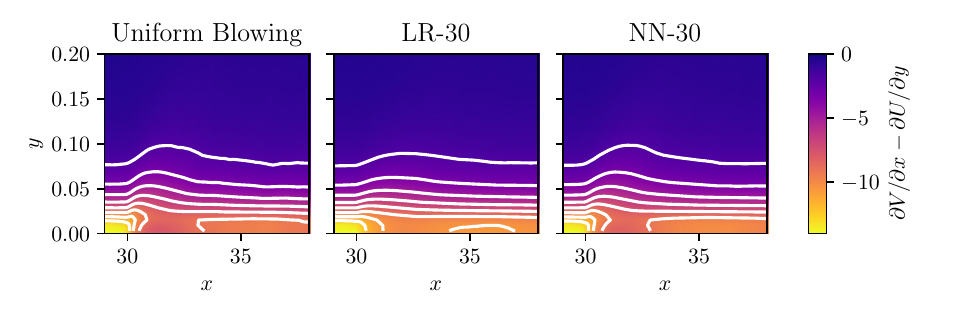}
    \caption{Mean spanwise vorticity around the beginning of the transpiration region.}
    \label{fig:mean_vorz}
\end{figure}

Figure \ref{fig:mean_vorz} depicts the spanwise vorticity for the relevant DNS cases with isocontours. In both coupled cases the mean vorticity at the wall near the beginning of the transpiration region is greater than in the uniform blowing case. The locations of the minimum near-wall vorticity coincide with the those of the observed peaks in the cooling effectiveness. 

\begin{figure}[htb]
    \centering
    \includegraphics[width=\textwidth]{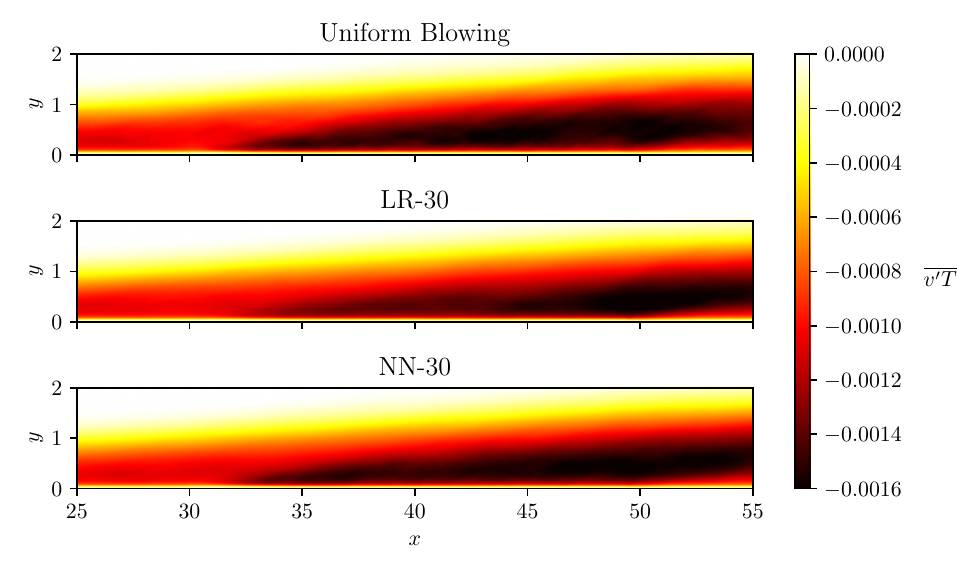}
    \caption{Mean turbulent transport of heat to the wall.}
    \label{fig:vT}
\end{figure}

The increase in near-wall vorticity is most apparent in the LR-30 case, which had the least pronounced peak in cooling effectiveness. This can be attributed to the low blowing ratio of this case at the beginning of the transpiration region; the injected coolant decreases the vorticity by reducing the near-wall velocity gradient, corresponding to a reduction in the turbulent transport of heat. Figure \ref{fig:vT} shows the mean turbulent transport of heat to the wall for the relevant DNS cases. 

Initially the transport is reduced in the coupled cases as compared to the uniform blowing case due to the more gradual introduction of coolant (reduced blowing ratio at the beginning of the transpiration region). Consequently there is initially less shear/vorticity generation, however as the rate of coolant injection increases further downstream the turbulent transport of heat eventually surpasses that of the uniform blowing case. Within the transpiration region this is offset by the increase in cooling ability associated with the additional coolant, however it reduces the effectiveness of the coolant film remaining after coolant injection ends.

\subsubsection{Heat advection and film accumulation}
The cooling effect in the boundary layer can be attributed to two effects: the decreasing temperature gradient at the wall due to the accumulation of a coolant film in the boundary layer, and the advection of heat away from the wall due to the non-zero wall-normal coolant injection velocity. It is of interest to see how both of these components are influenced by the introduced coupling.

\begin{figure}[htb]
    \centering
    \begin{subfigure}[b]{0.49\textwidth}
        \centering
        \includegraphics[width=\textwidth]{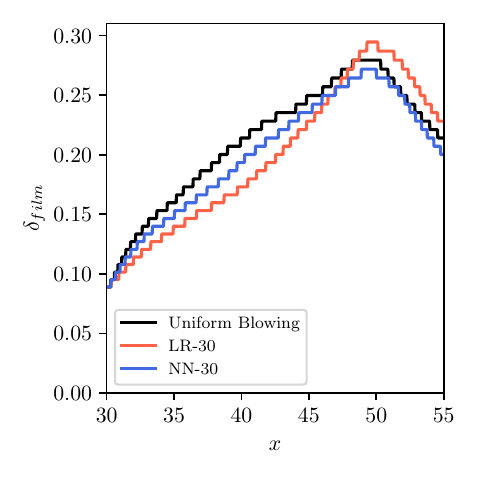}
        \caption{}
        \label{fig:film1D}
    \end{subfigure}
    \begin{subfigure}[b]{0.49\textwidth}
        \centering
        \includegraphics[width=\textwidth]{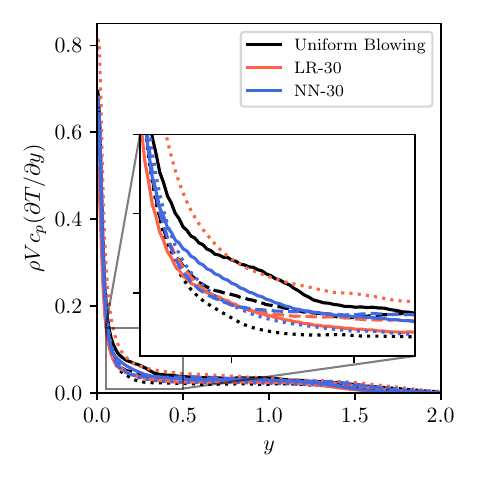}
        \caption{}
        \label{fig:eta_advec}
    \end{subfigure}
    \caption{Coolant film thickness (a) and effect of advection on heat transfer (b). In (b), linestyles correspond to different streamwise locations: $x=32.5$ (\full), $x=40$ (\dashed), $x=47.5$ (\dotted).}
\end{figure}

The first mechanism considered is the coolant film accumulation in the boundary layer, which modifies the boundary layer temperature profile, lowering the amount of heat flux entering the wall. To investigate this mechanism, the interface between the coolant film and the hot main flow was obtained by applying  an automatic interface identification algorithm based on a Fuzzy C-means clustering technique to the mean temperature field \cite{fan_detection_2019}. The streamwise development in mean film thickness found using this algorithm is plotted in Figure \ref{fig:film1D}.

As with the blowing ratio, the coolant film thickness in the coupled cases is lower than in the uniform blowing case at the beginning of the transpiration region. In the NN-30 case, this persists throughout the entirety of the transpiration region, however the film thickness in the LR-30 case surpasses that of the uniform blowing case at approximately $x=48$. In both cases, the overall cooling effectiveness as shown in Figure \ref{fig:mean_eta} exceeds that of the uniform blowing case prior to recovery of the coolant film from the coupling effects.

In order to gain a better understanding of this, it is necessary to look at the effect of advection on heat transfer throughout the transpiration region. To this end, the advection term $\overline{\rho}Vc_p(\partial\overline{T}/\partial y)$ for the coupled and uniform blowing cases at different streamwise locations is plotted in Figure \ref{fig:eta_advec}. This term is only significant very close to the wall where the turbulence is small; further out, turbulence dominates the heat transfer. In the uniform blowing case the advection term is strongest at the beginning of the transpiration region, decreasing further downstream due to the change in temperature gradient as the coolant film accumulates. However, in the coupled cases the blowing ratio and concomitantly the coolant injection velocity increase further downstream such that the advection term in the latter half of the transpiration region is greater for the coupled cases than for the uniform blowing case. This increase is large enough to compensate for the reduced film thickness, illustrating the importance of near-wall advective heat transfer on the downstream recovery of the cooling effectiveness in the coupled cases.

\subsection{Impact on turbulence}
\label{sec:results_turb}
\begin{figure}[htp]
    \centering
    \includegraphics[width=\textwidth]{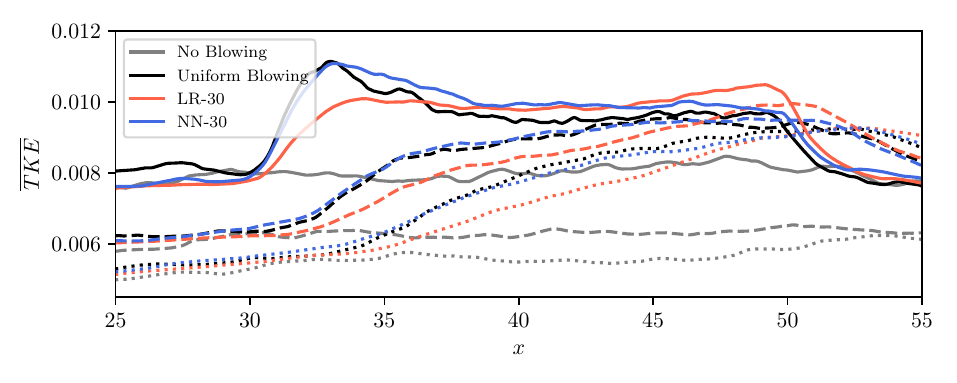}
    \caption{Time and spanwise-averaged TKE along the streamwise direction. Different linestyles correspond to different distances from the wall: $y=0.1$ (\full), $y=0.3$ (\dashed), $y=0.5$ (\dotted).}
    \label{fig:tke}
\end{figure}

In order to compare the turbulence in the different DNS cases, the time and spanwise-averaged turbulent kinetic energy (TKE) field was obtained for the different cases. Figure \ref{fig:tke} displays the streamwise distribution in TKE at three different wall-normal distances.


The introduction of coolant injection at the beginning of the transpiration region significantly increases the TKE of the flow, however the degree of this increase varies by the implemented coupling. The LR-30 case demonstrates the most gradual increase and a minimal peak at the onset of transpiration, which is in accordance with the mechanisms discussed in Section \ref{sec:turbheat_transport}. The TKE then slowly increases over the latter half of the transpiration region as the blowing ratio increases, resulting in additional turbulent production.

In contrast, while the NN-30 case also shows a slightly more gradual increase in near-wall TKE than its uniform blowing counterpart, it actually decreases over the latter half of the transpiration region despite the increasing blowing ratio. It is hypothesised that this is due to local coupling interactions modulating the turbulence (note that the observed decrease is only visible near the wall). In the NN-30 case the blowing ratio increases slowly; as the flow adjusts the attenuation of the pressure fluctuations due to the coupling is able to dominate, resulting in a gradual decrease in TKE. In contrast, the blowing ratio in the LR-30 case increases extremely rapidly downstream such that the increased shear and turbulent production dominates and the TKE increases downstream. This modulation of the near-wall turbulent structures due to the implemented coupling is explored further in Section \ref{sec:results-psd}.

\begin{figure}[ht]
    \centering
    \includegraphics[width=\textwidth]{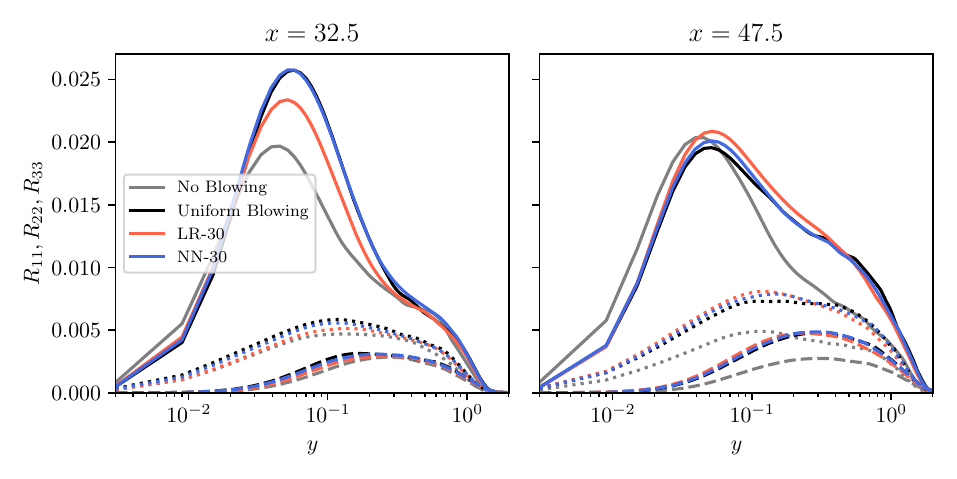}
    \caption{Reynolds normal stresses: $R_{11}$(\full), $R_{22}$(\dashed), $R_{33}$(\dotted).}
    \label{fig:mean_Re_stress}
\end{figure}

In addition to the TKE plots, the normal (diagonal component) Reynolds stresses are shown for all DNS cases in Figure \ref{fig:mean_Re_stress}. As the coupling is expected to impact the turbulence near the wall, a log scaling is applied to the x-axis to make differences in this region more easily visible.

The observed trends in the TKE of the flow discussed above are also present in all of the wall-normal profiles of the diagonal components of the Reynolds stresses, suggesting that the mechanism resulting in these changes to the turbulence of the flow is isotropic in nature, corroborating the theory that the increase in TKE is due to the shear produced by interactions between the increasing coolant injection and the main hot gas flow. The shear and resultant vortices due to this interaction are expected to produce velocity fluctuations in all directions, as observed.

\begin{figure}[htp]
    \centering
    \begin{subfigure}[b]{\textwidth}
        \centering
        \includegraphics[width=\textwidth]{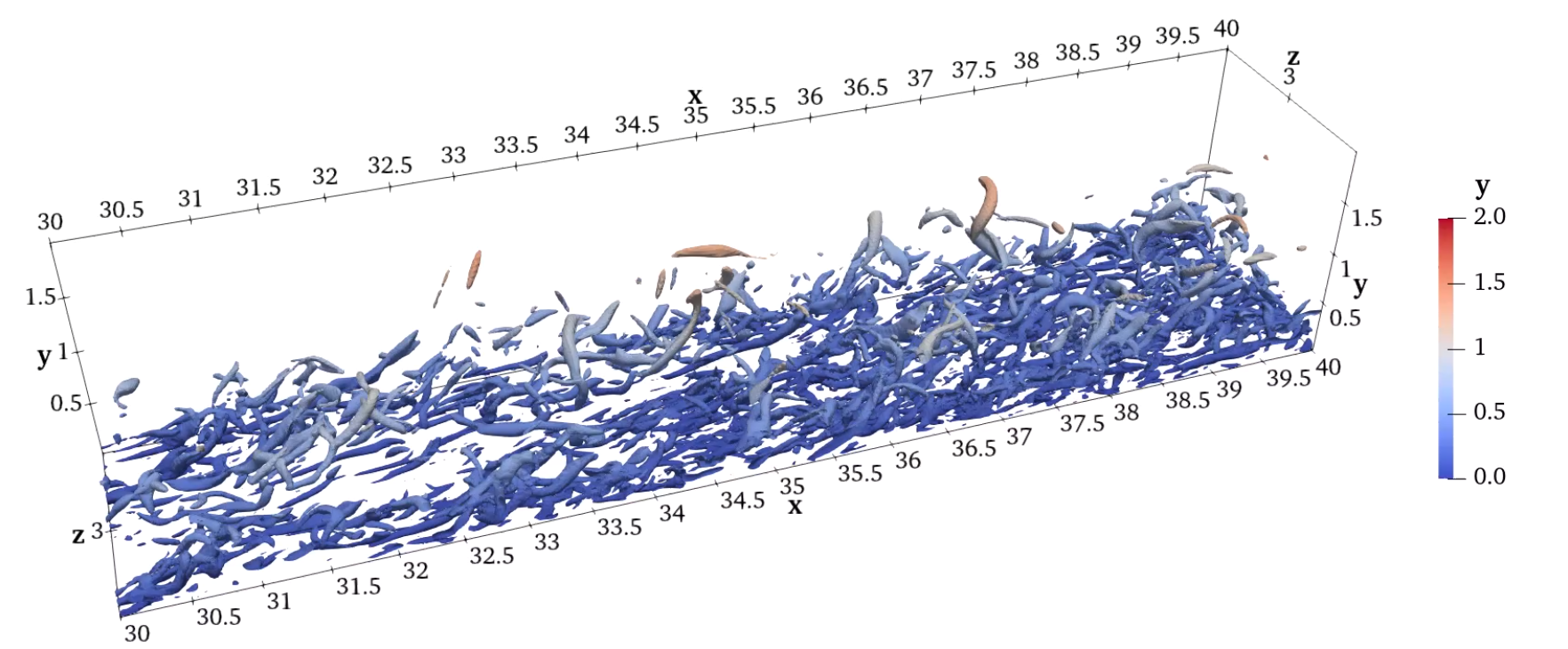}
        \caption{LR-30}
    \end{subfigure}
    \begin{subfigure}[b]{\textwidth}
        \centering
        \includegraphics[width=\textwidth]{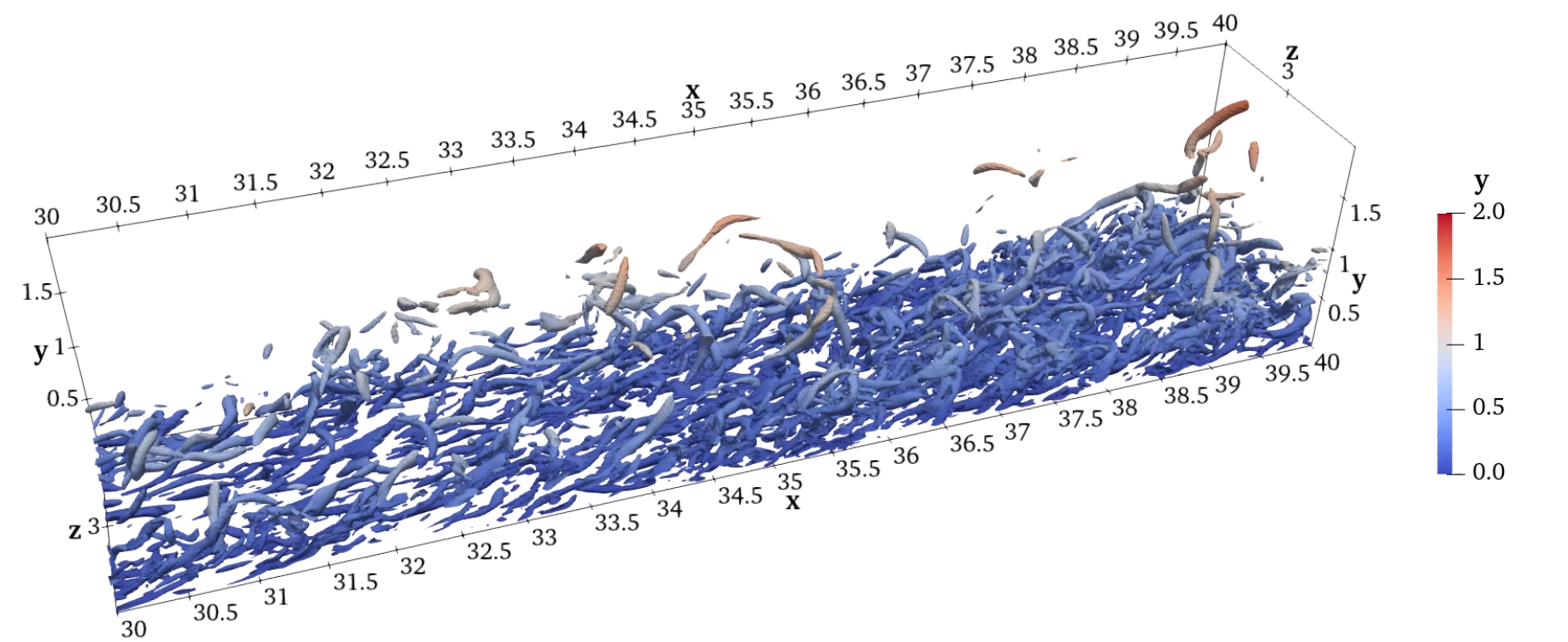}
        \caption{NN-30}
    \end{subfigure}
    \caption{Visualisation of large scale vortices in the first half of the transpiration region using isosurfaces of $\lambda_2=-2.$}
    \label{fig:turbvis_l2}
\end{figure}

An alternative method of analyzing the impact of the implemented coupling on the turbulence is via visualisation of the large scale vortical structures. This was done using the $\lambda_2$ criterion \cite{yao_toward_2018}. In this method, vortex cores are identified by locations in the flow where $\lambda_2<0$. The isosurfaces corresponding to $\lambda_2 = -2$ for the coupled cases are shown in Figure \ref{fig:turbvis_l2} in order to identify the location of strong vortices. Only half of the domain along the spanwise axis is shown, and the isosurfaces have been coloured according to the normal distance from the wall.

The observed trends, which were present in all extracted snapshots of each case, complement those found in the statistical analysis of the turbulence. At the beginning of the transpiration region there are fewer vortices in the LR-30 case than in the uniform blowing and NN-30 cases; however, the rapidly increasing blowing rate in the former introduces further turbulence such that the vortex density appears roughly similar for both shown cases further downstream, with the LR-30 case surpassing its counterparts at the very end of the transpiration region.

\subsection{Wall-pressure spectrum}
\label{sec:results-psd}
\begin{figure}[htb]
    \centering
    \includegraphics[width=0.98\textwidth]{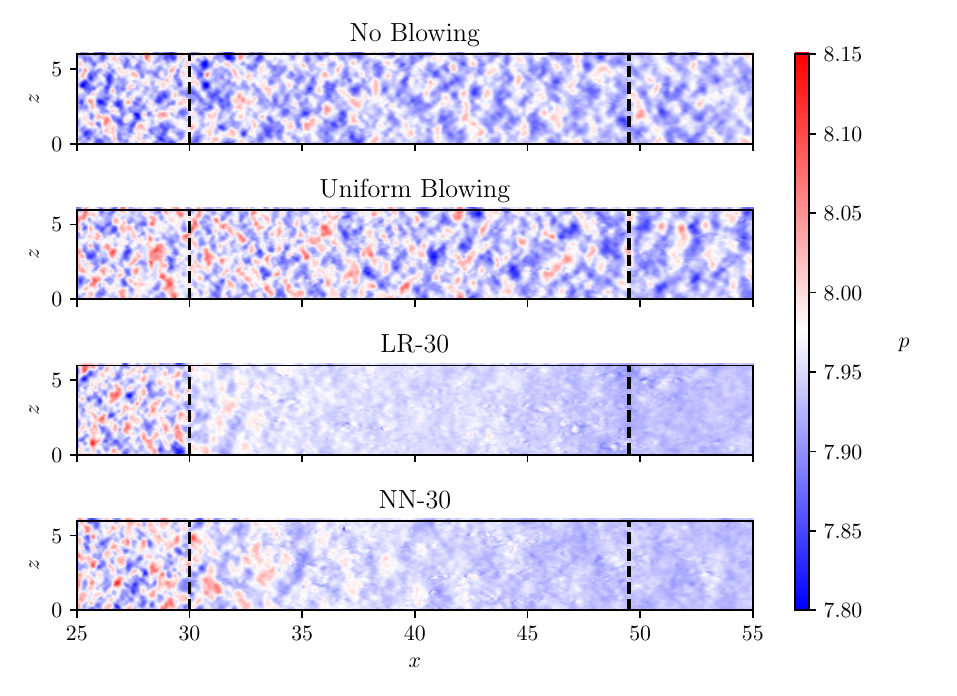}
    \caption{Instantaneous snapshots of large-scale pressure structures at the wall. The beginning and end of the transpiration region are demarcated by black dashed lines.}
    \label{fig:snapP_wall}
\end{figure}

A driving consideration of the present work was the turbulent pressure fluctuations that naturally occur at the wall in a turbulent boundary layer. Figure \ref{fig:snapP_wall} shows several instantaneous snapshots of the large scale pressure structures at the wall corresponding to the different cases considered. It is evident that coolant injection significantly modifies the near-wall pressure fluctuations.

In the uniform blowing case, the introduction of coolant injection clearly increases the pressure fluctuations both inside and upstream of the transpiration region due to the increase in produced shear and turbulence. The pressure structures near the beginning of the transpiration region are also noticeably compressed as a result of the impingement of the incoming main flow on the wall-normal coolant flow. Further downstream these structures stretch back out and are similar in size to their counterparts in the no blowing case, albeit with increased magnitude due to the additional turbulence.

The same structure compression and increase in fluctuations seen in the uniform blowing case is also observed upstream of the transpiration region in the coupled cases, however once transpiration begins these large pressure structures are rapidly attenuated due to the coupling between the injection velocity and the pressure fluctuations. The LR-30 cases disrupts the large scale structures very rapidly due to its increased sensitivity to changes in the pressure signal at the wall as discussed in Section \ref{sec:injection_dist}. These structures are also disrupted in the NN-30 case, however this occurs over a larger streamwise distance.

\begin{figure}[ht]
    \centering
    \includegraphics[width=\textwidth]{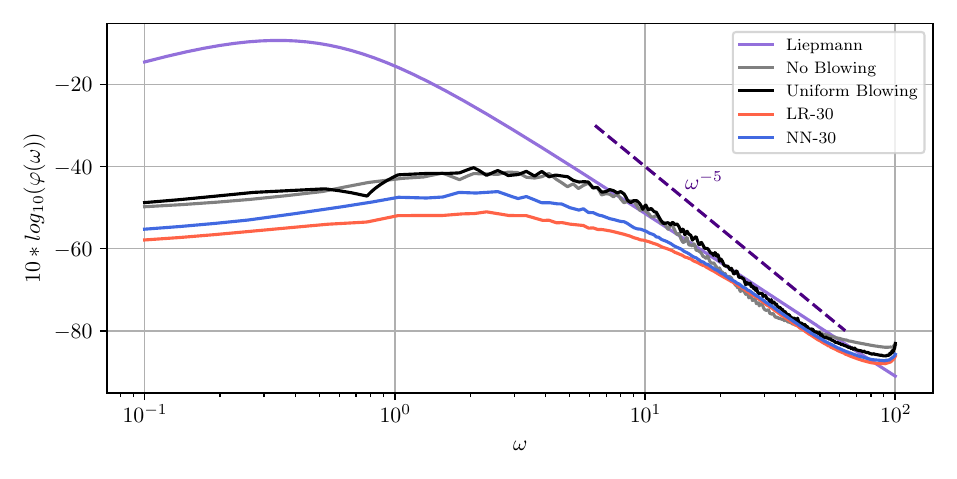}
    \caption{PSD of the DNS cases compared to that predicted by the analytical Liepmann model.}
    \label{fig:pressure_psd}
\end{figure}

It is important to note that the visible structures in Figure \ref{fig:snapP_wall} are only the largest, high energy-containing structures. In order to gain an understanding of how the implemented injection coupling interacts with the smallest pressure scales, the power spectral density (PSD) curves of the different cases within the transpiration region are shown in Figure \ref{fig:pressure_psd}.

The data output by the DNS cases was not sufficiently time-resolved to be able to directly calculate the frequency-based wall-pressure PSD within the frequency range of interest. Instead, a 2D Fast Fourier Transform (FFT) was used to obtain the PSD in the wavenumber domain, averaged over multiple snapshots. The analytical PSD was also calculated in the wavenumber domain, using the form of the Liepmann model \cite{liepmann_spectrum_1951} presented by Grasso et al.: 
\begin{equation}
    \varphi(\boldsymbol{k})^{TM}_{pp}=\frac{\rho_0^2\tau_w^2\overline{u'^2}\Lambda^5}{\pi \mu^2}\frac{k_1^2}{k + \gamma}\frac{8\zeta+9(k+\gamma)\Lambda\sqrt{\zeta}+3(k+\gamma)^2\Lambda^2}{\zeta^{(5/2)}\left((k+\gamma)\Lambda + \sqrt{\zeta})\right)^3}
\end{equation}
where $\zeta=1+k^2\Lambda^2$, $\rho_0=1$ is the fluid density in a quiescent medium (set to the free-stream density), $\Lambda=\delta_{99}$ is the integral length scale, and $\overline{{u'^2}}$ is the mean square fluctuation velocity component obtained via Reynolds decomposition. $k=\sqrt{k_1^2+k_3^2}$ is the planar wavenumber magnitude and  $\gamma=1/2\Lambda$ is the exponential wall function constant assumed during derivation.

The wavenumber signals were converted to the frequency domain using Taylor's hypothesis of frozen convection \cite{taylor_spectrum_1938}, which assumes a constant convection speed of the turbulent eddies in the flow direction \eqref{eq:freq_turb}. A mean convective speed of $U_c=0.8U_\infty$ was used here, as suggested by Grasso et al. \cite{grasso_analytical_2019}.

\begin{equation}
    \varphi_{pp}(\omega)=\frac{\int^{+\infty}_{-\infty}\varphi^{TM}_{pp}(k_c, k_3)\mathrm{d}k_3}{U_c}
    \label{eq:freq_turb}
\end{equation}

The Liepmann model performed surprisingly well compared to the no blowing case in the high-frequency region in spite of its assumptions that make it non-applicable, suggesting that the effects of the anisotropy of the turbulence and the flow temperature gradient can be adequately captured in this region by careful selection of the input flow parameters. As the frequency decreases the Liepmann PSD curve deviates from those obtained from the DNS cases, which begin to plateau, however this is a known limitation of the model attributed to an overlap between the inner and outer scaling parameters in this region \cite{grasso_analytical_2019}.

While the limited size and resolution of the DNS means that the PSD cannot be obtained for very low or high frequencies, two of the major regions identified by Goody \cite{goody_empirical_2004} are clearly present in both the no blowing and uniform blowing cases, these being the $\omega^{-5}$ trend at high frequencies, and the near plateau at middling frequencies (corresponding to a slope of $\omega^{-0.7}$). The increased near-wall turbulence in the uniform blowing case results in a slight increase of the associated pressure fluctuation PSD as anticipated, though due to the low blowing ratio the difference is relatively small. The impact is also somewhat offset by the formation of the coolant film, which pushes some of the structures responsible for the pressure fluctuations off of the wall.

By contrast, the PSD curves corresponding to the coupled cases differ significantly both from the analytical Liepmann model as well as the reference cases. The previously discussed attenuation of the large-scale pressure fluctuations, corresponding to lower frequencies, is clearly visible, with the effect being more significant in the LR-30 case than in the NN-30 case. These large structures account for a lot of the turbulent energy within the boundary layer and are expected to be particularly susceptible to decorrelation due to the variation in local injection velocity, causing them to be broken down and dissipated. 

Meanwhile, small-scale pressure fluctuations contain very little energy and cover very few pores, making them more robust against decorrelation. This is observed in Figure \ref{sec:results-psd}, where the PSD of the coupled cases is very close to that of the uniform blowing case at higher frequencies. While this too is an anticipated result, it is also important to keep in mind the inability of the neural network to capture and respond to the smaller-scale fluctuations. It is possible that some of the lack of attenuation of the high frequency structures in the neural network case is due to its difficulty responding to these structures, however the general trends discussed here are still expected to be relevant.

\section{Conclusions}
The present work sought to explore the coupling between coolant injection and turbulent pressure fluctuations in the boundary layer in transpiration cooling systems. This was motivated by a gap in the existing literature on the significance of this coupling and its implications with respect to the nature of the turbulence over the porous wall. To this end, an algorithm was developed to indirectly couple high-fidelity direct numerical simulations of flow in a turbulent boundary layer over a flatplate with a pore-network model through the use of a shallow convolutional neural network.

The impact of the coupling in the DNS cases coupled using linear expressions was much more pronounced as compared to those coupled using neural networks. This suggests that the incorporation of lateral flow between neighbouring pores has a modulating impact on the coupling and cannot be neglected.

In all coupled cases, these coupling effects resulted in a significant streamwise variation in the blowing ratio due to the high pressure over the wall at the beginning of the transpiration region. A corresponding reduction in cooling effectiveness and coolant film thickness was observed near the beginning of this region. All of these parameters eventually recovered as the blowing ratio increased. The importance of the advection of heat away from the wall in the recovery of the cooling effectiveness was highlighted.

The reduced blowing ratio at the beginning of the transpiration region resulted in reduced shear and, therefore, turbulent transport of heat to the wall, mitigating the initial peak and then decline in cooling effectiveness observed in the uniform blowing case. The streamwise variation in blowing ratio, however, resulted in increased turbulent production further downstream. In the neural network coupled cases, this effect was offset by an attenuation of the turbulence due to interactions of the resultant pressure fluctuations with the local coolant injection.

Finally, the influence of the imposed coupling on the pressure fluctuations at the porous wall was investigated. It was shown that the introduced coupling disrupts the high energy, large scale pressure structures while leaving the higher frequency fluctuations relatively untouched. In keeping with the trends observed in the other results, the cases coupled using the linear expressions attenuated the large scale fluctuations nearly instantaneously due to the high sensitivity of the coupling to small changes in pressure, while the fluctuations were able to persist slightly further downstream in the neural network cases.



\section*{Acknowledgments}
This research was enabled in part by support provided by Sharcnet, Scinet, and the Digital Research Alliance of Canada through the RAC program. 
We would like to thank Mr. Nicholas Christopher for the reuse of some comparative simulation data. SH and JPH acknowledge the support of the Natural Sciences and Engineering Research Council of Canada.

 \bibliographystyle{elsarticle-num} 
 \bibliography{references,ref_JPH}





\end{document}